\documentclass[sigconf]{acmart}
\usepackage{amsmath}
\usepackage{hyperref}
\usepackage{fancyvrb} 
\usepackage{tcolorbox}
\usepackage{stfloats}
\usepackage{float}
\usepackage{tabularx}
\usepackage{multirow}

\renewcommand\footnotetextcopyrightpermission[1]{} 

\AtBeginDocument{%
  }

\acmConference[KDD '24]{the 30th ACM SIGKDD Conference on Knowledge Discovery
and Data Mining}{August 25-29,
  2024}{Barcelona}

\begin{document}

\title{Revisiting the Solution of Meta KDD Cup 2024: CRAG}

\author{Jie Ouyang}
\affiliation{%
  \institution{State Key Laboratory of Cognitive Intelligence,  University of Science and Technology of China}
  \city{Hefei}
  \state{Anhui}
  \country{China}
}
\email{ouyang_jie@mail.ustc.edu.cn}

\author{Yucong Luo}
\affiliation{%
  \institution{State Key Laboratory of Cognitive Intelligence,  University of Science and Technology of China}
  \city{Hefei}
  \state{Anhui}
  \country{China}
}
\email{prime666@mail.ustc.edu.cn}

\author{Mingyue Cheng}
\authornote{Mingyue Cheng is the corresponding author.}
\orcid{0000-0001-9873-7681}
\affiliation{%
  \institution{State Key Laboratory of Cognitive Intelligence,  University of Science and Technology of China}
  \city{Hefei}
  \state{Anhui}
  \country{China}
}
\email{mycheng@ustc.edu.cn}

\author{Daoyu Wang}
\affiliation{%
  \institution{State Key Laboratory of Cognitive Intelligence,  University of Science and Technology of China}
  \city{Hefei}
  \state{Anhui}
  \country{China}
}
\email{wdy030428@mail.ustc.edu.cn}

\author{Shuo Yu}
\affiliation{%
  \institution{State Key Laboratory of Cognitive Intelligence,  University of Science and Technology of China}
  \city{Hefei}
  \state{Anhui}
  \country{China}
}
\email{yu12345@mail.ustc.edu.cn}

\author{Qi Liu}
\affiliation{%
  \institution{State Key Laboratory of Cognitive Intelligence,  University of Science and Technology of China}
  \city{Hefei}
  \state{Anhui}
  \country{China}
}
\email{qiliuql@ustc.edu.cn}

\author{Enhong Chen}
\affiliation{%
  \institution{State Key Laboratory of Cognitive Intelligence,  University of Science and Technology of China}
  \city{Hefei}
  \state{Anhui}
  \country{China}
}
\email{cheneh@ustc.edu.cn}

\begin{abstract}
  This paper presents the solution of our team APEX in the Meta KDD CUP 2024: CRAG Comprehensive RAG Benchmark Challenge. The CRAG benchmark addresses the limitations of existing QA benchmarks in evaluating the diverse and dynamic challenges faced by Retrieval-Augmented Generation (RAG) systems. It provides a more comprehensive assessment of RAG performance and contributes to advancing research in this field. We propose a routing-based domain and dynamic adaptive RAG pipeline, which performs specific processing for the diverse and dynamic nature of the question in all three stages: retrieval, augmentation, and generation. Our method achieved superior performance on CRAG and ranked 2nd for Task 2\&3 on the final competition leaderboard. Our implementation is available at this link: \href{https://github.com/USTCAGI/CRAG-in-KDD-Cup2024}{https://github.com/USTCAGI/CRAG-in-KDD-Cup2024}.
\end{abstract}

\ccsdesc[300]{Information systems~Information retrieval}

\keywords{Retrieval-Augmented Generation, Large Language Model}



\maketitle

\section{Introduction}

Large Language Models (LLMs) have revolutionized the landscape of Natural Language Processing (NLP) tasks \cite{Cheng2024TowardsPE, Luo2023UnlockingTP, Jiang2023ReformulatingSR}, particularly in question answering (QA). Despite advances in LLMs, hallucination remains a significant challenge, particularly for dynamic facts and information about less prominent entities.

Retrieval-Augmented Generation (RAG) \cite{lewis2020retrieval} has recently emerged as a promising solution to mitigate LLMs' knowledge deficiencies. Given a question, a RAG system queries external sources to retrieve relevant information and subsequently provides grounded answers. Despite its potential, RAG continues to face numerous challenges, including the selection of the most relevant information, the reduction of question answering latency, and the synthesis of information to address complex questions.

To bridge this gap, Meta introduced the Comprehensive RAG Benchmark (CRAG) \cite{yang2024crag}, a factual question answering benchmark of 4,409 question-answer pairs and Mock APIs to simulate web and Knowledge Graph (KG) search, and hosted the KDD CUP 2024 Challenge.

\subsection{Dataset Description}
The CRAG contains two parts of data: the QA pairs and the content for retrieval.

\textbf{QA pairs}. The CRAG dataset contains a rich set of 4,409 QA pairs covering five domains: finance, sports, music, movie, and open domain, and eight types of questions. For the KDD CUP 2024 Challenge, the benchmark data were splited into three sets with similar distributions: validation, public test, and private test at 30\%, 30\%, and 40\%, respectively. In total, 2,706 examples from validation and public test sets were shared.

The dataset also reflects varied entity popularity from popular to long-tail entities, and temporal spans ranging from seconds to years. Given the temporal nature of many questions, each question-answer pair is accompanied by an additional field denoted as "query time." This temporal marker ensures the reliability and uniqueness of the answers within their specific temporal context.


\textbf{Content for retrieval}. The CRAG dataset incorporates two types of content for retrieval to simulate a practical scenario for RAG: web search and knowledge graph (KG) search. This encompasses up to 50 full HTML pages for each question, retrieved from a real-world search engine, as well as Mock KGs containing 2.6 million entities. Additionally, CRAG provides Mock APIs to simulate retrieval from a wide range of available information sources.

\begin{itemize}
    \item \textbf{Web search results} Each retrieved web search result comprises five fields: \textit{page name}, \textit{page url}, \textit{page snippet}, \textit{page last modified} and \textit{page result}. See Table \ref{tab:websearch} in Appendix \ref{app:web} for an example.
    \item \textbf{Mock KGs} A Knowledge Graph containing 2.6 million entities, which is accessed through Mock API.
    \item \textbf{Mock APIs} APIs for retrieving structured data from Mock KGs with predefined parameters, which are categorized by domain and output in JSON format. See Example in Appendix \ref{app:mock}
.    
\end{itemize}

\subsection{Task Desription}
This challenge comprises three tasks designed to improve question-answering (QA) systems.

\textbf{TASK 1:} 
The organizers provide 5 web pages per question, potentially containing relevant information. The objective is to measure the systems' capability to identify and condense this information into accurate answers.

\textbf{TASK 2:}
This task introduces mock APIs to access information from underlying mock Knowledge Graphs (KGs), with structured data possibly related to the questions. Participants use mock APIs, inputting parameters derived from the questions, to retrieve relevant data for answer formulation. The evaluation focuses on the systems' ability to query structured data and integrate information from various sources into comprehensive answers.

\textbf{TASK 3:}
The third task increases complexity by providing 50 web pages and mock API access for each question, encountering both relevant information and noise. It assesses the systems' skill in selecting the most important data from a larger set, reflecting the challenges of real-world information retrieval and integration.

Each task builds upon the previous, steering participants toward developing sophisticated end-to-end RAG systems. This challenge showcases the potential of RAG technology in navigating and making sense of extensive information repositories, setting the stage for future AI research and development breakthroughs.

\section{Methodlogy}

\begin{figure*}[ht]
  \centering
  \includegraphics[width=\linewidth]{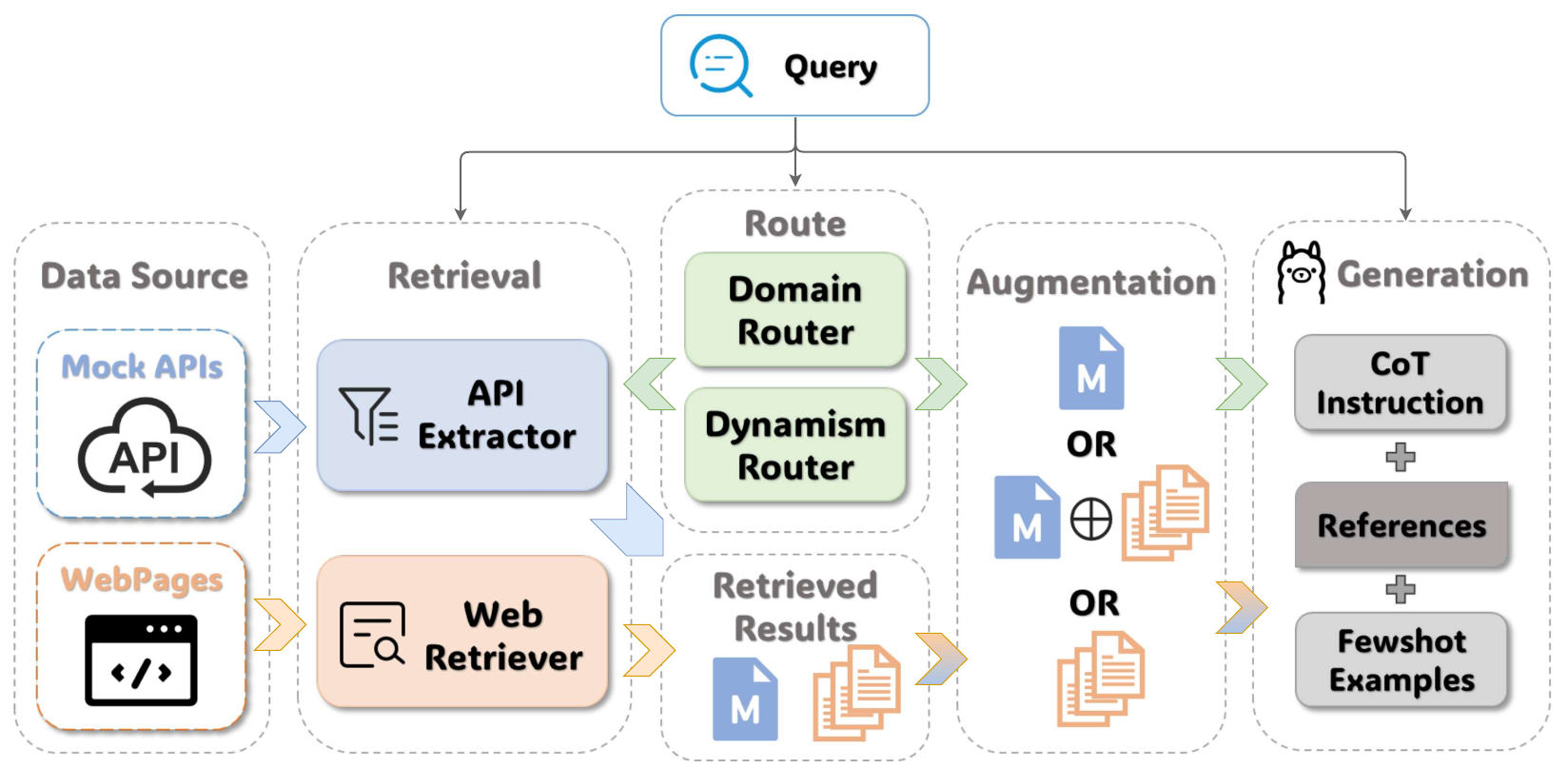}
  \caption{The overall pipeline of our solution.}
  \label{fig:pipeline}
\end{figure*}

Our approach, akin to most Retrieval-Augmented Generation (RAG) systems, comprises three primary phases: retrieval, augmentation and generation. In all phases, we implement routing mechanisms to address diverse query types. Figure \ref{fig:pipeline} illustrates the pipeline of our solution, while the remainder of this section details the three main phases and the routers employed in this challenge.

\subsection{Router}
Routing is a crucial component of RAG systems, especially in real-world QA scenarios. In practical applications, RAG systems frequently incorporate multiple data sources. In the CRAG Challenge, we have three distinct data sources: Web Pages, Mock KGs, and Mock APIs. The diversity of questions requires routing queries to different data sources, individually or in combination. Even within a single data source, such as Mock APIs, the question-specific selection of appropriate APIs is crucial. Furthermore, we can tailor prompt templates based on the nature of the question or route questions to different post-processing components.

In response to the specific characteristics of the questions in the CRAG Challenge, we designed two specialized routers: the Domain Router and the Dynamism Router. These routers are designed to efficiently navigate the complex landscape of multisource information retrieval and question-specific processing in our RAG system.

\textbf{Domain Router}. Domain router is fundamentally a classifier, more specifically, a sentence classifier. Given a query, the domain router assigns a specific domain from a predefined set: finance, sports, music, movie, and open. Based on the assigned domain, the workflow is then routed to the corresponding path.

We utilize \textit{Llama3-8B-Instruct} \cite{meta2024llama3} as our base model and enhance it with a classification head (Multilayer Perceptron, MLP) for domain classification. The 8B model inherently demonstrates a robust capability to comprehend the domain of queries. We randomly split the CRAG dataset into training, validation, and test sets with a ratio of 8:1:1. Based on this split, we performed a simple LORA (Low-Rank Adaptation) \cite{Hu2021LoRALA} fine-tuning to adapt to the distribution of the CRAG dataset. This approach facilitates the development of a high-quality classifier with minimal additional training.

The trained Domain router is employed at multiple stages within the system. During the retrieval phase, the Domain router is primarily used to select appropriate APIs. Following the retrieval of Web Pages and APIs, it is further applied for selective fusion of the retrieved knowledge. In the generation phase, we first customize the prompt templates based on the domain. Subsequently, after the model completes its generation, the Domain is also utilized for corresponding post-processing.

\textbf{Dynamic Router}. Analogous to the Domain router, the Dynamism router is also a sentence classifier. Given a question, the Dynamism router assigns a specific Dynamism, specifically one of: static, slow-changing, fast-changing, or real-time. The specific training methodology for the Dynamism router is congruent with that of the Domain Router, and thus will not be recapitulated here.

Due to the inherent limitation of Large Language Models in updating their internal knowledge, they are prone to providing outdated answers to dynamic questions. Even when employing external knowledge through RAG, LLMs can readily generate hallucinations as most data sources are static, unless real-time APIs are utilized. In the absence of real-time APIs, the more rapidly a question changes over time, the more susceptible LLMs are to hallucinations.

To attenuate hallucinations arising from dynamic questions, we implemented the Dynamism router for post-processing. In scenarios where real-time APIs are inaccessible, we excluded certain real-time questions and those that change rapidly over time.

\subsection{Retrieval}

As mentioned above, the CRAG dataset encompasses three types of content for retrieval: Web Pages, Mock KGs, and Mock APIs. For our final solution, we utilize two of these content types: Web Pages and Mock APIs.

\begin{figure}[h]
  \centering
  \includegraphics[width=\linewidth]{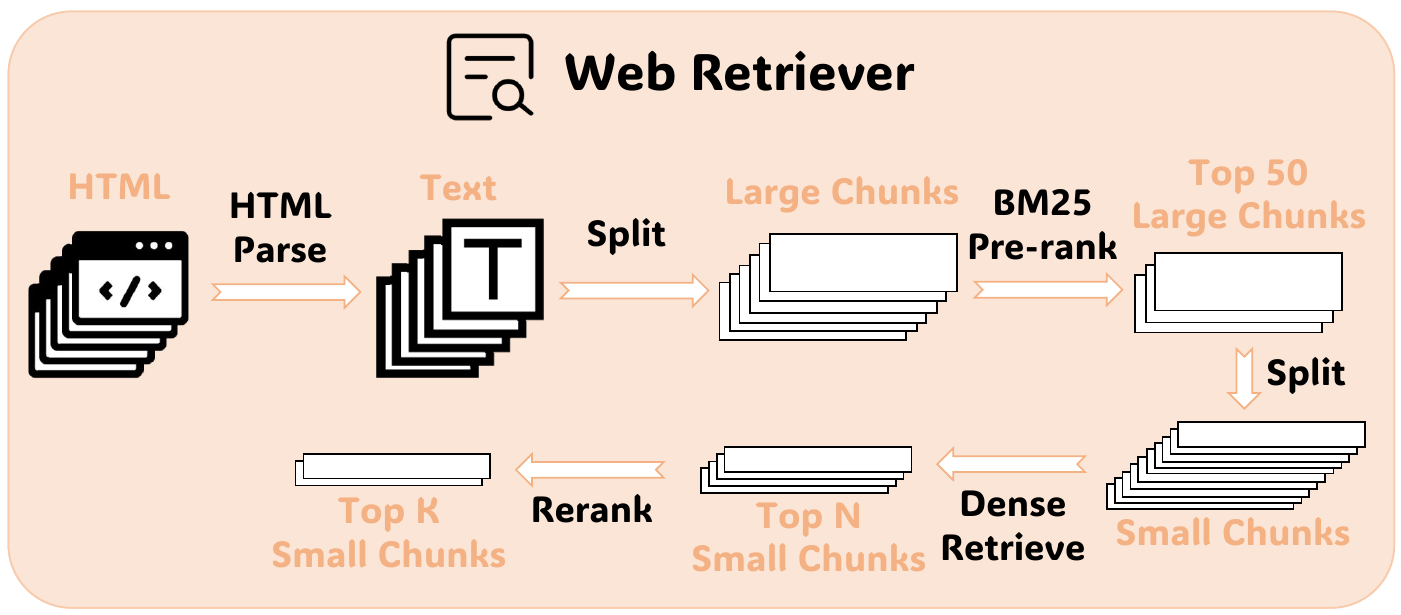}
  \caption{The pipeline of Web Retriever.}
  \label{fig:retrieval pipeline}
\end{figure}

\subsubsection{\textbf{Web Pages}}
Web Pages are available for all 3 tasks of the challenge. For Task 1\&2, 5 Web Pages per question are provided, each potentially containing relevant information. Task 3 increases complexity by offering 50 Web Pages for each question, presenting both pertinent information and noise.

To enhance our QA system, we need to extract useful and relevant information from web search results. The primary process for retrieving web content is illustrated in Figure \ref{fig:retrieval pipeline}. 

\begin{enumerate}
    \item \textbf{HTML Parsing}: Structured HTML is often unnecessarily verbose and contains substantial extraneous information that can impede subsequent segmentation operations. Therefore, it is crucial to first convert this structured format into natural language text that is more amenable to processing by Large Language Models. We conducted experiments with various HTML parsing methods, including BeautifulSoup, Newspaper, Markdownify, and several others. After evaluating both parsing efficiency and quality, we ultimately selected Newspaper. See \ref{app:html} for more details about experiment results.
    \item \textbf{Pre-ranking} (Task 3 only): For Task 3, fine-grained processing 50 Web Pages would be excessively time-consuming. Therefore, we initially filter out an appropriate amount of relevant text before ranking. Specifically, we segment all the text from the Web Pages into chunks of 1024 tokens (calculated based on tokens rather than characters). For these segmented text chunks, we use BM25 \cite{robertson2009probabilistic} to select the top 50 most relevant text blocks.
    \item \textbf{Ranking}: In the ranking phase, we further refine the pre-ranked text blocks. For task 1\&2, the text blocks are the 5 raw plain text extracted from HTML. The text blocks are segmented into smaller chunks, each comprising 256 tokens. We then transform these 256-token chunks into embeddings utilizing the \textit{bge-m3} \cite{chen2024bge} model. Finally, we calculate the cosine similarity  to select top 10 relevant chunks.  
    \item \textbf{Re-ranking}: We utilized \textit{bge-m3-v2-reranker} \cite{flagembedding2024reranker} to re-rank the aforementioned 10 relevant chunks, ultimately selecting the top 5 segments.
\end{enumerate}

\begin{figure}[h]
  \centering
  \includegraphics[width=\linewidth]{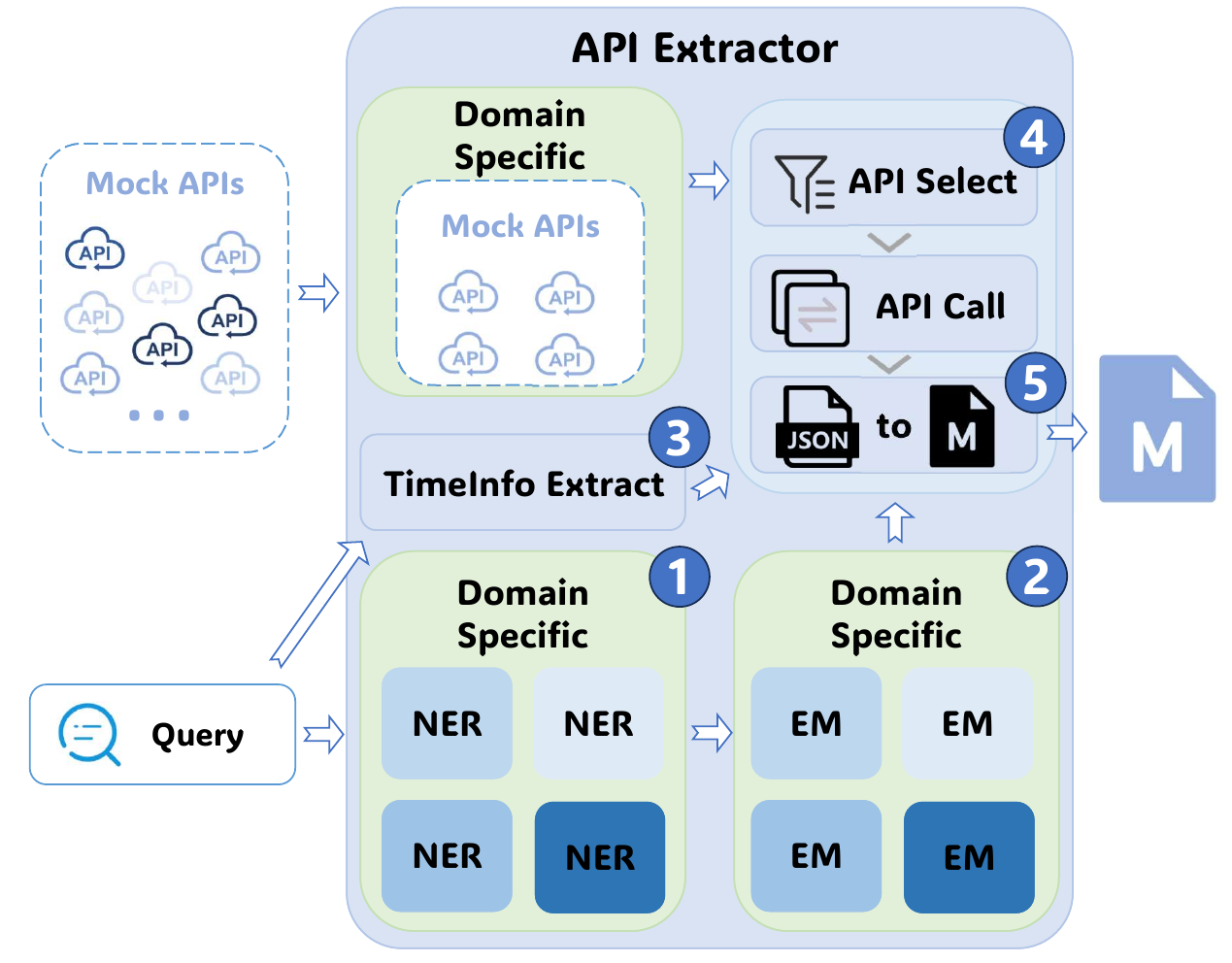}
  \caption{The pipeline of API Extractor.}
  \label{fig:api pipeline}
\end{figure}

\subsubsection{\textbf{Mock APIs}} 
A total of 38 Mock APIs were provided for tasks 2\&3. As mentioned above, these Mock APIs can be categorized into five distinct domains, with no overlap between the APIs of different domains. Naturally, we designed separate workflows for each domain using a Domain Router. However, the overarching process flow of the workflows across all domains remains consistent, as shown in Figure \ref{fig:api pipeline}:
\begin{enumerate}
    \item \textbf{Named Entity Recognition (NER)}: We directly utilize \textit{Llama3-70B-Instruct} \cite{meta2024llama3} to identify and classify named entities in the question into predefined categories, such as movie names and artist names. Specific prompts are presented in the Appendix \ref{app:ner}.
    \item \textbf{Entity Match}: Matching extracted entities with API input parameters. Taking finance as an example, the input parameter for finance API is typically the ticker symbol, while user questions often contain full company names. We need to convert the company names to their corresponding ticker symbols. We first perform exact matching, requiring the extracted entity to be exactly the same as the input parameter. If no match is found, we then use BM25 to select the most similar one.
    \item \textbf{Time Information Extraction}: In addition to entities, numerous API inputs incorporate temporal information, requiring the extraction of relevant time points or intervals from user inputs. Notably, temporal information is often dependent with query time, as illustrated by terms such as "yesterday." In such cases, we must determine the specific time point based on the query time through relative time computation. We first use regular expressions to match certain time and date-related terms. Once matched, we use two Python packages (pytz and datetime) to calculate the Absolute Time. If no matches are found, the current time is used by default.
    \item \textbf{API Select}: Each domain comprises numerous APIs, not all of which are inherently relevant to a user's query. We have manually designed a set of rules to select APIs that correspond to the given question. To minimize the risk of overfitting the rules to the training set, we implemented constraints on the rule design process, prioritizing simplicity and robustness.
    \item \textbf{Json to Markdown}: The JSON output from APIs, while structured and machine-readable, may not be optimal for large language models (LLMs) to process efficiently. Converting this JSON data into a more LLM-friendly Markdown format can enhance the model's ability to understand and utilize the information.
\end{enumerate}

\subsection{Augmentation}

We employ input-layer integration for generation augmentation, which combines retrieved
information/documents with the original input/query and jointly passes them to the generator. In contrast to common input-layer integration, we do not utilize all retrieved documents. For different domains, we select specific data sources and integrate them to construct the final reference.

For the open domain, since we did not employ a Mock API, we exclusively utilized web search results. For the movie and music domains, where most queries are relatively static or evolve slowly, results retrieved from both web pages and mock APIs remain relevant. Therefore, we chose to integrate these two sources. For the sports and finance domains, which involve numerous real-time and fast-changing queries, we exclusively used Mock APIs to ensure the timeliness and relevance of the retrieved information.

\subsection{Generation}

In the generation phase, we employed two widely-used methodologies: Chain-of-Thought (CoT) reasoning and In-context Learning. After generation, we performed a simple post-processing procedure on the generated results based on the Domain and Dynamism Routers.

\subsubsection{\textbf{Chain of Thought}}

Chain-of-Thought (CoT) \cite{wei2022chain} enhances the reasoning process of language models by prompting them to articulate intermediate steps in problem-solving. This approach not only enhances the model's ability to handle complex tasks but also significantly reduces hallucinations.

\subsubsection{\textbf{In-context Learning}}

We improve the model's ability to recognize invalid questions, particularly those based on false premises, through In-context Learning. We develop adaptive few-shot examples \cite{10.5555/3495724.3495883}, selecting two of the most representative invalid question samples for each domain and elucidating the reasons for their invalidity. Using the sports domain as an example, our few-shot samples are as follows:

\begin{tcolorbox}
[title=\textbf{Few-shot Example \uppercase\expandafter{\romannumeral1}}]
What's the latest score for OKC's game today?
\tcblower
\textit{There is no game for OKC today.}
\end{tcolorbox}

\begin{tcolorbox}
[title=\textbf{Few-shot Example \uppercase\expandafter{\romannumeral2}}]
How many times has Curry won the NBA dunk contest?
\tcblower
\textit{Steph Curry has never participated in the NBA dunk contest.}
\end{tcolorbox}




\subsubsection{\textbf{Post-processing}}

Before finalizing the results, we implement basic post-processing strategies. Based on the question domain and volatility, we assign "I don't know" responses to queries susceptible to hallucination. For domains lacking real-time API access, specifically open, movie, and music categories, we designated "I don't know" answers to fast-changing and real-time questions. Furthermore, due to the model's limited mathematical computation capabilities, we implement same processing for questions requiring numerical calculations, such as those involving the term "average." Although these approach may be deemed simplistic, it demonstrated efficacy in significantly reducing hallucinations.

    
    
    
    


\section{Experiments}
\begin{table*}[ht]
\caption{Overall Preformance of our solutions on all 3 Tasks.}
\setlength{\tabcolsep}{12pt} 
\renewcommand{\arraystretch}{1} 
\begin{tabular}{lrrrr}
\toprule
\multicolumn{1}{l}{\textbf{}} & \multicolumn{1}{l}{\textbf{Score(\%)}} & \multicolumn{1}{l}{\textbf{Accuracy(\%)}} & \multicolumn{1}{l}{\textbf{Hallucination(\%)}} & \multicolumn{1}{l}{\textbf{Missing(\%)}} \\
\midrule
\textbf{LLM Only}             & -7.29                         & 28.01                            & 35.30                                 & 36.69                           \\
\textbf{Direct RAG}         & -6.78                         & 34.79                            & 41.58                                 & \textbf{23.63}                  \\
\midrule
\textbf{Task 1}               & 11.82                         & 29.98                            & 18.16                                 & 51.86                           \\
\textbf{Task 2}               & 31.22                         & 46.75                            & \textbf{15.54}                        & 37.71                           \\
\textbf{Task 3}               & \textbf{31.66}                & \textbf{48.21}                   & 16.56                                 & 35.23                           \\
\bottomrule
\end{tabular}
\label{tab:overall}
\end{table*}

\begin{table*}[ht]
\caption{Ablation Study for Major Strategies Employed in the System.}
\setlength{\tabcolsep}{10pt} 
\renewcommand{\arraystretch}{1} 
\begin{tabular}{clrrrrr}
\toprule
\multicolumn{2}{l}{} & \multicolumn{1}{l}{\textbf{Score(\%)}} & \multicolumn{1}{l}{\textbf{Accuracy(\%)}} & \multicolumn{1}{l}{\textbf{Hallucination(\%)}} & \multicolumn{1}{l}{\textbf{Missing(\%)}} & \multicolumn{1}{l}{\textbf{Time Cost(s)}}\\
\midrule
\multirow{7}{*}{\textbf{Task 2}} & \textbf{w/o Rerank} & 29.17 & 43.54 & 14.37 & 42.09 & - \\
& \textbf{w/o EntityMatch} & 21.44 & 32.31 & 10.87 & 56.82 & - \\
& \textbf{w/o TimeInfoExtract} & 18.45 & 28.45 & \textbf{9.99} & 61.56 & - \\
& \textbf{w/o Fewshot\&CoT} & 25.53 & 52.08 & 26.55 & \textbf{21.37} & - \\
& \textbf{w/o Fewshot} & 27.13 & 51.35 & 24.22 & 24.43 & - \\
& \textbf{w/o CoT} & 28.52 & \textbf{53.32} & 24.80 & 21.88 & - \\
& \textbf{Ours} & \textbf{31.22} & 46.75 & 15.54 & 37.71 & - \\
\midrule
\multirow{2}{*}{\textbf{Task 3}} & \textbf{w/o Prerank} & 29.53 & 44.34 & \textbf{14.81} & 40.85 & 68.17 \\
& \textbf{Ours} & \textbf{31.66} & \textbf{48.21} & 16.56 & \textbf{35.23} & \textbf{5.96} \\
\bottomrule
\end{tabular}
\label{tab:ablation}
\end{table*}

In this section, we present our main results and ablation studies for
some crucial components.

We did not employ a strategy of fine-tuning the LLM; instead, we used the LLM in a zero-shot setting. According to the rules set by the organizers, we used \textit{Llama3-70B-Instruct} \cite{meta2024llama3} for all our LLMs. For the embedding model, we used \textit{BAAI/bge-m3}, and for the rerank model, we used \textit{BAAI/bge-m3-v2-reranker}. In the 1371 public test cases officially released for this round, we compared the following baselines: LLM only (using the LLM without retrieving references) and straightforward RAG (the baseline provided by the organizers, using straightforward RAG solutions).

\subsection{Metrics and Evaluation}


In line with CRAG Benchmark, we conduct a model-based automatic evaluation for our experiment. Automatic evaluation employs rule-based matching and GPT-4 \cite{achiam2023gpt} assessment to check answer correctness. It will assign three scores: correct (1 point), missing (0 points), and incorrect (-1 point). The final score for a given RAG system is computed as the average score across all examples in the evaluation set.

\subsection{Overall Performance}

Comparing our solutions to the RAG Baseline, we observe significant advantages in performance across all three tasks. In Table \ref{tab:overall}, our approach showcases notable improvements in accuracy and information retention. Specifically, when contrasted with the RAG Baseline, our solutions demonstrate superior results with reduced hallucination rates and enhanced information completeness. Task 2 and Task 3, in particular, exhibit substantial enhancements in accuracy and reduced hallucination percentages, highlighting the effectiveness of our proposed methodologies in addressing these key metrics.

\subsection{Ablation Study}

Table \ref{tab:ablation} presents the ablation study for major strategies employed in our solution. 

During the retrieval phase, we implemented several strategies, including pre-ranking, re-ranking, Entity Match, and Time Information Extraction. Ablation studies revealed that pre-ranking and re-ranking marginally reduce performance, while Entity Match and Time Information Extraction significantly decrease performance. Both pre-ranking and re-ranking significantly contribute to the improvement of retrieval quality. Pre-ranking enhances retrieval performance by proactively filtering out a significant amount of noise, while re-ranking ensures the accuracy of retrieval results through more refined and granular sorting. The enhancement in retrieval quality ultimately translates into an increase in answer accuracy. The absence of pre-ranking and re-ranking demonstrably leads to a substantial decrease in the accuracy of the final answers. Furthermore, pre-ranking significantly enhances retrieval effiency and reduces the retrieval time. Entity Matching and Time Information Extraction form the basis of using MOCK APIs. They ensure the accuracy of API call parameters, which is crucially linked to the overall performance. The absence of either component can result in a significant performance decline.

During the generation phase, we employed 2 main components: domain specific fewshot examples and Chain of Thought prompt. Both aforementioned components led to a substantial reduction in hallucinations. The combined implementation of these components yielded a reduction in hallucinations of up to 71\%, consequently resulting in a 22\% increase in the final score. The experimental results demonstrate the effectiveness and necessity of the two components.

\section{Perspectives}

Our method presents a robust and versatile framework for addressing a wide range of dynamic and complex real-world problems. This approach, however, also opens up several avenues for further investigation.

\begin{itemize}
    \item \textbf{Model Cognitive Ability Assessment} Most conventional QA evaluation methods primarily focus on accuracy, neglecting the impact of hallucinations. Models should be aware of their knowledge boundaries, discerning what they should and should not answer. CRAG incorporates hallucinations into evaluation metrics, but its settings lack sufficient justification. Responding "I don't know" to all questions can yield a satisfactory score. Exploring the assessment of models' cognitive abilities using methodologies for evaluating human cognition is a promising research direction.
    \item \textbf{API Integration and Scalability}. In real-world scenarios, where extensive API usage is common, our manually designed matching rules are likely to prove inadequate. The development of a more universal method for selecting and calling APIs, as well as processing the returned results, represents a promising avenue for future research.
    \item \textbf{Handling Dynamic Information}. For questions that involve information that changes dynamically over time, simply refusing to answer is merely a basic solution. Future research should focus on exploring methods to acquire the most up-to-date knowledge and determine the timeliness of information. This is crucial to avoid hallucinations caused by outdated knowledge and to ensure the system provides accurate, current information. Developing techniques for real-time information retrieval and verification, as well as implementing mechanisms to assess the reliability and currency of data sources, are key areas for investigation.
\end{itemize}

\section{Conclusion}
In this paper, we introduce our solution for the Meta KDD CUP 2024: CRAG Comprehensive RAG Benchmark. We adopt a classic RAG framework with two specific routers. In the retrieval phase, we demonstrated the process of obtaining high-quality information from various data sources and utilizing the Domain Router for information filtering. In the augmentation phase, we employed the Domain Router in a similar manner for information aggregation based on domain characteristics. Finally, in the generation phase, we implemented two methods to significantly improve the model's accuracy and reduce hallucinations, while further mitigating hallucinations through post-processing based on the question's domain and dynamic nature.

Our approach offers a viable pathway for addressing the diverse and dynamic challenges encountered in real-world scenarios. Nevertheless, our method has certain limitations. We have identified several inherent issues in the current methodology and provided our insights and reflections on the specific problems related to our approach. Ultimately, we anticipate that this study will make a modest contribution to the broader RAG and LLM communities.

\begin{acks}
This research was supported by grants from the National Key Research and Development Program of China (Grant No. 2021YFF0901000), the National Natural Science Foundation of China (No. 62337001) and the Fundamental Research Funds for the Central Universities.
\end{acks}

\bibliographystyle{ACM-Reference-Format}
\bibliography{sample-base}


\begin{thebibliography}{13}


\ifx \showCODEN    \undefined \def \showCODEN     #1{\unskip}     \fi
\ifx \showDOI      \undefined \def \showDOI       #1{#1}\fi
\ifx \showISBNx    \undefined \def \showISBNx     #1{\unskip}     \fi
\ifx \showISBNxiii \undefined \def \showISBNxiii  #1{\unskip}     \fi
\ifx \showISSN     \undefined \def \showISSN      #1{\unskip}     \fi
\ifx \showLCCN     \undefined \def \showLCCN      #1{\unskip}     \fi
\ifx \shownote     \undefined \def \shownote      #1{#1}          \fi
\ifx \showarticletitle \undefined \def \showarticletitle #1{#1}   \fi
\ifx \showURL      \undefined \def \showURL       {\relax}        \fi
\providecommand\bibfield[2]{#2}
\providecommand\bibinfo[2]{#2}
\providecommand\natexlab[1]{#1}
\providecommand\showeprint[2][]{arXiv:#2}

\bibitem[Achiam et~al\mbox{.}(2023)]%
        {achiam2023gpt}
\bibfield{author}{\bibinfo{person}{Josh Achiam}, \bibinfo{person}{Steven Adler}, \bibinfo{person}{Sandhini Agarwal}, \bibinfo{person}{Lama Ahmad}, \bibinfo{person}{Ilge Akkaya}, \bibinfo{person}{Florencia~Leoni Aleman}, \bibinfo{person}{Diogo Almeida}, \bibinfo{person}{Janko Altenschmidt}, \bibinfo{person}{Sam Altman}, \bibinfo{person}{Shyamal Anadkat}, {et~al\mbox{.}}} \bibinfo{year}{2023}\natexlab{}.
\newblock \showarticletitle{Gpt-4 technical report}.
\newblock \bibinfo{journal}{\emph{arXiv preprint arXiv:2303.08774}} (\bibinfo{year}{2023}).
\newblock


\bibitem[AI(2024)]%
        {meta2024llama3}
\bibfield{author}{\bibinfo{person}{Meta AI}.} \bibinfo{year}{2024}\natexlab{}.
\newblock \showarticletitle{Meta LLaMA 3}.
\newblock \bibinfo{journal}{\emph{Meta AI Blog}} (\bibinfo{year}{2024}).
\newblock
\urldef\tempurl%
\url{https://ai.meta.com/blog/meta-llama-3/}
\showURL{%
\tempurl}


\bibitem[Brown et~al\mbox{.}(2020)]%
        {10.5555/3495724.3495883}
\bibfield{author}{\bibinfo{person}{Tom~B. Brown}, \bibinfo{person}{Benjamin Mann}, \bibinfo{person}{Nick Ryder}, \bibinfo{person}{Melanie Subbiah}, \bibinfo{person}{Jared Kaplan}, \bibinfo{person}{Prafulla Dhariwal}, \bibinfo{person}{Arvind Neelakantan}, \bibinfo{person}{Pranav Shyam}, \bibinfo{person}{Girish Sastry}, \bibinfo{person}{Amanda Askell}, \bibinfo{person}{Sandhini Agarwal}, \bibinfo{person}{Ariel Herbert-Voss}, \bibinfo{person}{Gretchen Krueger}, \bibinfo{person}{Tom Henighan}, \bibinfo{person}{Rewon Child}, \bibinfo{person}{Aditya Ramesh}, \bibinfo{person}{Daniel~M. Ziegler}, \bibinfo{person}{Jeffrey Wu}, \bibinfo{person}{Clemens Winter}, \bibinfo{person}{Christopher Hesse}, \bibinfo{person}{Mark Chen}, \bibinfo{person}{Eric Sigler}, \bibinfo{person}{Mateusz Litwin}, \bibinfo{person}{Scott Gray}, \bibinfo{person}{Benjamin Chess}, \bibinfo{person}{Jack Clark}, \bibinfo{person}{Christopher Berner}, \bibinfo{person}{Sam McCandlish}, \bibinfo{person}{Alec Radford}, \bibinfo{person}{Ilya Sutskever},
  {and} \bibinfo{person}{Dario Amodei}.} \bibinfo{year}{2020}\natexlab{}.
\newblock \showarticletitle{Language models are few-shot learners}. In \bibinfo{booktitle}{\emph{Proceedings of the 34th International Conference on Neural Information Processing Systems}} (Vancouver, BC, Canada) \emph{(\bibinfo{series}{NIPS '20})}. \bibinfo{publisher}{Curran Associates Inc.}, \bibinfo{address}{Red Hook, NY, USA}, Article \bibinfo{articleno}{159}, \bibinfo{numpages}{25}~pages.
\newblock
\showISBNx{9781713829546}


\bibitem[Chen et~al\mbox{.}(2024)]%
        {chen2024bge}
\bibfield{author}{\bibinfo{person}{Jianlv Chen}, \bibinfo{person}{Shitao Xiao}, \bibinfo{person}{Peitian Zhang}, \bibinfo{person}{Kun Luo}, \bibinfo{person}{Defu Lian}, {and} \bibinfo{person}{Zheng Liu}.} \bibinfo{year}{2024}\natexlab{}.
\newblock \showarticletitle{Bge m3-embedding: Multi-lingual, multi-functionality, multi-granularity text embeddings through self-knowledge distillation}.
\newblock \bibinfo{journal}{\emph{arXiv preprint arXiv:2402.03216}} (\bibinfo{year}{2024}).
\newblock


\bibitem[Cheng et~al\mbox{.}(2024)]%
        {Cheng2024TowardsPE}
\bibfield{author}{\bibinfo{person}{Mingyue Cheng}, \bibinfo{person}{Hao Zhang}, \bibinfo{person}{Jiqian Yang}, \bibinfo{person}{Qi Liu}, \bibinfo{person}{Li Li}, \bibinfo{person}{Xin Huang}, \bibinfo{person}{Liwei Song}, \bibinfo{person}{Zhi Li}, \bibinfo{person}{Zhenya Huang}, {and} \bibinfo{person}{Enhong Chen}.} \bibinfo{year}{2024}\natexlab{}.
\newblock \showarticletitle{Towards Personalized Evaluation of Large Language Models with An Anonymous Crowd-Sourcing Platform}.
\newblock \bibinfo{journal}{\emph{Companion Proceedings of the ACM on Web Conference 2024}} (\bibinfo{year}{2024}).
\newblock
\urldef\tempurl%
\url{https://api.semanticscholar.org/CorpusID:268379217}
\showURL{%
\tempurl}


\bibitem[FlagOpen(2024)]%
        {flagembedding2024reranker}
\bibfield{author}{\bibinfo{person}{FlagOpen}.} \bibinfo{year}{2024}\natexlab{}.
\newblock \bibinfo{title}{FlagEmbedding Reranker}.
\newblock \bibinfo{howpublished}{\url{https://github.com/FlagOpen/FlagEmbedding/tree/master/FlagEmbedding/reranker}}.
\newblock


\bibitem[Hu et~al\mbox{.}(2021)]%
        {Hu2021LoRALA}
\bibfield{author}{\bibinfo{person}{J.~Edward Hu}, \bibinfo{person}{Yelong Shen}, \bibinfo{person}{Phillip Wallis}, \bibinfo{person}{Zeyuan Allen-Zhu}, \bibinfo{person}{Yuanzhi Li}, \bibinfo{person}{Shean Wang}, {and} \bibinfo{person}{Weizhu Chen}.} \bibinfo{year}{2021}\natexlab{}.
\newblock \showarticletitle{LoRA: Low-Rank Adaptation of Large Language Models}.
\newblock \bibinfo{journal}{\emph{ArXiv}}  \bibinfo{volume}{abs/2106.09685} (\bibinfo{year}{2021}).
\newblock
\urldef\tempurl%
\url{https://api.semanticscholar.org/CorpusID:235458009}
\showURL{%
\tempurl}


\bibitem[Jiang et~al\mbox{.}(2023)]%
        {Jiang2023ReformulatingSR}
\bibfield{author}{\bibinfo{person}{Junzhe Jiang}, \bibinfo{person}{Shang Qu}, \bibinfo{person}{Mingyue Cheng}, {and} \bibinfo{person}{Qi Liu}.} \bibinfo{year}{2023}\natexlab{}.
\newblock \showarticletitle{Reformulating Sequential Recommendation: Learning Dynamic User Interest with Content-enriched Language Modeling}.
\newblock \bibinfo{journal}{\emph{ArXiv}}  \bibinfo{volume}{abs/2309.10435} (\bibinfo{year}{2023}).
\newblock
\urldef\tempurl%
\url{https://api.semanticscholar.org/CorpusID:262054865}
\showURL{%
\tempurl}


\bibitem[Lewis et~al\mbox{.}(2020)]%
        {lewis2020retrieval}
\bibfield{author}{\bibinfo{person}{Patrick Lewis}, \bibinfo{person}{Ethan Perez}, \bibinfo{person}{Aleksandra Piktus}, \bibinfo{person}{Fabio Petroni}, \bibinfo{person}{Vladimir Karpukhin}, \bibinfo{person}{Naman Goyal}, \bibinfo{person}{Heinrich K{\"u}ttler}, \bibinfo{person}{Mike Lewis}, \bibinfo{person}{Wen-tau Yih}, \bibinfo{person}{Tim Rockt{\"a}schel}, {et~al\mbox{.}}} \bibinfo{year}{2020}\natexlab{}.
\newblock \showarticletitle{Retrieval-augmented generation for knowledge-intensive nlp tasks}.
\newblock \bibinfo{journal}{\emph{Advances in Neural Information Processing Systems}}  \bibinfo{volume}{33} (\bibinfo{year}{2020}), \bibinfo{pages}{9459--9474}.
\newblock


\bibitem[Luo et~al\mbox{.}(2023)]%
        {Luo2023UnlockingTP}
\bibfield{author}{\bibinfo{person}{Yucong Luo}, \bibinfo{person}{Mingyue Cheng}, \bibinfo{person}{Hao Zhang}, \bibinfo{person}{Junyu Lu}, \bibinfo{person}{Qi Liu}, {and} \bibinfo{person}{Enhong Chen}.} \bibinfo{year}{2023}\natexlab{}.
\newblock \showarticletitle{Unlocking the Potential of Large Language Models for Explainable Recommendations}.
\newblock \bibinfo{journal}{\emph{ArXiv}}  \bibinfo{volume}{abs/2312.15661} (\bibinfo{year}{2023}).
\newblock
\urldef\tempurl%
\url{https://api.semanticscholar.org/CorpusID:266551720}
\showURL{%
\tempurl}


\bibitem[Robertson et~al\mbox{.}(2009)]%
        {robertson2009probabilistic}
\bibfield{author}{\bibinfo{person}{Stephen Robertson}, \bibinfo{person}{Hugo Zaragoza}, {et~al\mbox{.}}} \bibinfo{year}{2009}\natexlab{}.
\newblock \showarticletitle{The probabilistic relevance framework: BM25 and beyond}.
\newblock \bibinfo{journal}{\emph{Foundations and Trends{\textregistered} in Information Retrieval}} \bibinfo{volume}{3}, \bibinfo{number}{4} (\bibinfo{year}{2009}), \bibinfo{pages}{333--389}.
\newblock


\bibitem[Wei et~al\mbox{.}(2022)]%
        {wei2022chain}
\bibfield{author}{\bibinfo{person}{Jason Wei}, \bibinfo{person}{Xuezhi Wang}, \bibinfo{person}{Dale Schuurmans}, \bibinfo{person}{Maarten Bosma}, \bibinfo{person}{Fei Xia}, \bibinfo{person}{Ed Chi}, \bibinfo{person}{Quoc~V Le}, \bibinfo{person}{Denny Zhou}, {et~al\mbox{.}}} \bibinfo{year}{2022}\natexlab{}.
\newblock \showarticletitle{Chain-of-thought prompting elicits reasoning in large language models}.
\newblock \bibinfo{journal}{\emph{Advances in neural information processing systems}}  \bibinfo{volume}{35} (\bibinfo{year}{2022}), \bibinfo{pages}{24824--24837}.
\newblock


\bibitem[Yang et~al\mbox{.}(2024)]%
        {yang2024crag}
\bibfield{author}{\bibinfo{person}{Xiao Yang}, \bibinfo{person}{Kai Sun}, \bibinfo{person}{Hao Xin}, \bibinfo{person}{Yushi Sun}, \bibinfo{person}{Nikita Bhalla}, \bibinfo{person}{Xiangsen Chen}, \bibinfo{person}{Sajal Choudhary}, \bibinfo{person}{Rongze~Daniel Gui}, \bibinfo{person}{Ziran~Will Jiang}, \bibinfo{person}{Ziyu Jiang}, {et~al\mbox{.}}} \bibinfo{year}{2024}\natexlab{}.
\newblock \showarticletitle{CRAG--Comprehensive RAG Benchmark}.
\newblock \bibinfo{journal}{\emph{arXiv preprint arXiv:2406.04744}} (\bibinfo{year}{2024}).
\newblock


\end{thebibliography}

\appendix

\section{Appendix}

\subsection{An Example of Web Search Results.}
\label{app:web}

\begin{table}[H]
\caption{An Example of Web Search Results.}
\label{tab:websearch}
\begin{tabularx}{\linewidth}{lX}
\hline
Key & Value \\
\hline
"page name" & "Microsoft Office 2019 - Wikipedia" \\
"page url" & "https://en.wikipedia.org/wiki/\newline
Microsoft\_Office\_2019" \\
"page snippet" & "For Office 2013 and 2016, various editions containing the client apps were available in both Click-To-Run..." \\
"page last modified" & "Tue, 27 Feb 2024 22:55:55 GMT" \\
"html page" & "<!DOCTYPE html>\newline
<html class=...>\newline
<head>\newline
...\newline
<title>Microsoft Office 2019 - Wikipedia</title>..." \\
\hline
\end{tabularx}
\end{table}

\subsection{An Example of Mock APIs.}
\label{app:mock}

Most mock APIs typically take entities as input and output entity-related information in JSON format. The following is an example of an API in the economic domain:
\begin{tcolorbox}[title=\texttt{get\_detailed\_price\_history}]
\textbf{Description:} The function returns the past 5 days' history of 1-minute Stock price, starting from 09:30:00 EST to 15:59:00 EST.

\textbf{Input:}
\begin{itemize}
    \item \texttt{ticker\_name}: \texttt{str}, upper case
\end{itemize}

\textbf{Output:}
\begin{itemize}
    \item Past 5 days' 1-minute price history: \texttt{json}
\end{itemize}
\end{tcolorbox}

\subsection{Experiment Results of HTML Parsing Methods}
\label{app:html}

\begin{table}[h]
\caption{Experiment Results of HTML Parsing Methods}
\begin{tabular}{c|rrr}
\hline
\textbf{Method}        & \textbf{Time Cost(s)} & \textbf{Success Rate(\%)} & \textbf{Score(\%)} \\ \hline
\textbf{beautifulsoup} & 0.31            & \textbf{100.0}              & 5.11               \\
\textbf{boilerpy3}     & \textbf{0.25}         & 97.2                      & 9.42               \\
\textbf{trafilatura}   & 0.64                  & 96.4                      & 10.14              \\
\textbf{newspaper}     & 1.96                  & 98.8                & \textbf{11.82}     \\
\textbf{markdownify}   & 3.65                  & \textbf{100.0}              & 10.94        \\ \hline
\end{tabular}
\end{table}

\subsection{Prompts Used for Name Entity Recognition}
\label{app:ner}

\begin{tcolorbox}
[title=\textbf{Music NER Prompt}]
{\itshape 
Please identify and list all the named entities present in the following question about music instead answering it, categorizing them appropriately (e.g., persons, song, band) Your answer should be short and concise in 50 words.

~

Format your response as follows: For each entity, provide the name followed by its category in parentheses. Categories include persons, songs and bands. Ensure that your response is clearly structured and easy to read.

~

Question: "\{query\}"

~

Output only the named entities present in the question. Do not include any other information. If there are no named entities in the question, please provide an empty response.

~

Expected Output Format:

a name of a person in the sentence (person)

a name of a song in the sentence (song)

a name of a band in the sentence (band);

~

Every entity should be in a new line and be in the format of "entity\_name (entity\_category)"
}
\end{tcolorbox}

\begin{tcolorbox}
[title=\textbf{Sports NER Prompt}]
{\itshape 
Please identify and list all the named entities present in the following question about sports instead answering it, categorizing them appropriately (e.g., nba team, soccer team, nba player, soccer player) Your answer should be short and concise in 50 words.

~

Format your response as follows: For each entity, provide the name followed by its category in parentheses. Categories include nba teams, soccer teams, nba players, soccer players. Ensure that your response is clearly structured and easy to read.

~

Question: "\{query\}"

~

Output only the named entities present in the question. Do not include any other information. If there are no named entities in the question, please provide an empty response.

~

Expected Output Format:

a name of a nba team in the sentence  (nba team)

a name of a soccer team in the sentence  (soccer team)

a name of a nba player in the sentence  (nba player)

a name of a soccer player in the sentence (soccer player);

~

Every entity should be in a new line and be in the format of "entity\_name (entity\_category)"
}
\end{tcolorbox}

\begin{tcolorbox}
[title=\textbf{Movie NER Prompt}]
{\itshape 
Please identify and list all the named entities present in the following question about movie instead answering it, categorizing them appropriately (e.g., person, movie) Your answer should be short and concise in 50 words.

~

Format your response as follows: For each entity, provide the name followed by its category in parentheses. Categories include persons, and movies. Ensure that your response is clearly structured and easy to read.

~

Question: "\{query\}"

~

Output only the named entities present in the question. Do not include any other information. If there are no named entities in the question, please provide an empty response.

~

Expected Output Format:

a name of a person in the sentence (person)

a name of a movie in the sentence (movie)

~

Every entity should be in a new line and be in the format of "entity\_name (entity\_category)"
}
\end{tcolorbox}


\begin{tcolorbox}
[title=\textbf{Finance NER Prompt}]
{\itshape 
Please identify and list all the named entities present in the following question about finance instead answering it, categorizing them appropriately (e.g., company, ticker symbol) Your answer should be short and concise in 50 words.

~

Format your response as follows: For each entity, provide the name followed by its category in parentheses. Categories include company, and symbol(which means the ticker symbol of a company). Ensure that your response is clearly structured and easy to read.

~

Question: "\{query\}"

~

Output only the named entities present in the question. Do not include any other information. If there are no named entities in the question, please provide an empty response.

~

Expected Output Format:

a name of a company in the sentence  (company)

a name of a ticker symbol in the sentence (symbol);

~

Every entity should be in a new line and be in the format of "entity\_name (entity\_category)"
}
\end{tcolorbox}

\end{document}